\documentstyle[12pt]{article}
\begin{document}
\hoffset-10mm
\voffset-20mm
\vskip4cm\begin{center}
{\LARGE\bf
Square of General Relativity
 }\footnote{{\em A talk at the First Australasian Conference
 on General Relativity and Gravitation, Adelaide, 
 February 12--17, 1996}, to be published in the Proceedings.}\\ \vskip1cm
{\bf
D.V. Gal'tsov}\footnote{Email: galtsov@grg.phys.msu.su} 
\\
\normalsize  Department of Theoretical Physics,  Moscow State University,\\ 
\normalsize Moscow 119899, {\bf Russia}\\ 
\vskip1cm

{\bf Abstract}\end{center}
\begin{quote}
We consider dilaton--axion gravity interacting with $p\;\, U(1)$ vectors 
($p=6$ corresponding to $N=4$ supergravity) in four--dimensional spacetime 
admitting a non--null Killing vector field. It is argued that this theory 
exibits features of a ``square'' of vacuum General Relativity. In 
the three--dimensional formulation it is equivalent to a gravity coupled 
$\sigma$--model with the $(4+2p)$--dimensional target space 
$SO(2,2+p)/(SO(2)\times SO(2+p))$. K\"ahler coordinates are introduced on the 
target manifold generalising Ernst potentials  of General Relativity. The 
corresponding K\"ahler potential is found to be equal to the logarithm of the 
product of the four--dimensional metric component $g_{00}$ in the Einstein 
frame and the dilaton factor, independently on presence of vector fields. 
The K\"ahler potential is invariant under exchange of the Ernst potential and 
the complex axidilaton field, while it undergoes holomorphic/antiholomorphic 
transformations under general target space isometries. The ``square'' property 
is also manifest in the two--dimensional reduction of the theory as a matrix 
generalization of the Kramer--Neugebauer map.
   
\vskip5mm
\noindent
PASC number(s): 97.60.Lf, 04.60.+n, 11.17.+y
\end{quote}
\newpage
 
Supergravity is often called  a ``square root'' of General Relativity.
Indeed, a supersymmetric extention of the Poincare algebra is
reminiscent of the Dirac's procedure of obtaining a spin 1/2 wave
equation from the scalar wave equation. A bosonic sector of {\it extended}
supergravities, apart from the graviton, contains scalar and vector fields.
One of the most interesting bosonic structures is suggested by the $N=4$
supergravity, which attracted much attention recently in connection
with ``stringy'' black holes \cite{gi, rk4, gk, cv}. 
Here we want to discuss the relationship between this theory (often called
also dilaton--axion gravity) and vacuum General Relativity and to show
that, in a certain sense, it can be viewed as a ``square'' of the latter.

As far as {\it stationary} solutions, like
black holes, are concerned (more generally, space--times possessing
a non--null Killing vector field), Einstein theory may be 
reformulated as a three--dimensional gravity coupled non--linear
$\sigma$--model \cite{nk}. This theory admits a concise representation in 
terms of the Ernst potential $\epsilon$ \cite{er, ex},
which may be regarded as a complex coordinate on a one--dimensional K\"ahler 
manifold $SU(1,1)/U(1)$. Similar representation holds for the dilaton--axion 
$\sigma$--model. The K\"ahler potential for the target space of the 
three--dimensional $N=4$ supergravity turns out to be equal to the logarithm 
of the product of potentials of the vacuum gravity and the dilaton--axion
system independently on presence of vector fields. For vacuum
gravity this potential is simply given by the logarithm of four--dimensional
$g_{00}$. A dilaton exponential plays similar role in the dilaton--axion 
$\sigma$--model possessing the same $SU(1,1)/U(1)$ structure. 
A square of this coset generates the target space of the pure 
dilaton--axion gravity, which is enlarged to the manifold 
$SO(2,2+p)/(SO(2)\times SO(2+p))$ when vector fields are included. However, 
the K\"ahler potential still preserves its value given by the
pure dilaton--axion gravity. This gives rise to various similarities between 
classical solutions of the Einstein--Maxwell--dilaton--axion theory
and vacuum Einstein equations. One can say that $N=4$ supergravity exibits
at the same time features of the square root and the square of General
Relativity. Apart from this somewhat phylosophical implication,
the K\"ahler representation of the stationary $N=4$ supergravity
turns out to be very useful in identifying hidden symmetries and 
generating classical solutions.

Let us recall first the Ernst formulation of the stationary vacuum 
Einstein equations \cite{er}. Assuming the spacetime to admit a 
timelike (in an essential region) Killing vector field, one parametrises
the four--dimensional metric through the standard Kaluza--Klein
ansatz
\begin{equation}
ds^2=g_{\mu\nu}dx^\mu dx^\nu=f(dt-\omega_idx^i)^2-\frac{1}{f}h_{ij}dx^idx^j,
\end{equation}
where the three--space metric $h_{ij},\;(i, j=1, 2, 3)$, the rotation 
one--form $\omega_i $ and the three--dimensional conformal factor $f$ 
depend on the three--space coordinates $x^i$ only. 
Then it can be shown that the vanishing of the mixed components of 
the Ricci tensor $R^i_0$ implies the existence of the NUT--potential
$\chi$ replacing a rotation one--form $\omega=\omega_i dx^i$ via
dualization \cite{iw}
\begin{equation}  
d\chi=-f^2*d\omega,
\end{equation}
where an asterisk stands for the three--dimensional Hodge dual.
Together $f$ and $\chi$ parametrize a two--dimensional target
manifold of the $\sigma$--model resulting from dimensional reduction. 
A K\"ahler representation is achieved by introducing the following
linear combination, the (vacuum) Ernst potential,
\begin{equation}  
\epsilon =if-\chi,
\end{equation}
(more frequently used potential differs from this by $i$). 
The corresponding equations of motion define a harmonic map from the 
three--dimensional space ${x^i}$ to the target manifold 
endowed with the metric
\begin{equation}  
dl^2=2 G_{\epsilon {\bar \epsilon}}d\epsilon d{\bar \epsilon}=
-2\frac{d\epsilon d{\bar \epsilon}}{(\bar \epsilon-\epsilon)^2}.
\end{equation}

The K\"ahler metric $G_{\epsilon {\bar \epsilon}}$ is generated by
the K\"ahler potential $K$
\begin{equation}  
G_{\epsilon {\bar \epsilon}}=\partial_\epsilon \partial_{\bar\epsilon} 
K(\epsilon,\, {\bar\epsilon}),
\end{equation}
for which one obtains the following simple expression
\begin{equation}  
K=-\ln V,\quad V={\rm Im} \epsilon=f.
\end{equation}
Thus the K\"ahler potential for any stationary solution of the 
vacuum Einstein equation is directly related to the $g_{00}$
component of the four--dimensional metric.

Ernst potential act as a source in the three--dimensional Einstein 
equations for the metric $h_{ij}$
\begin{equation}  
{\cal R}_{ij}=\frac{1}{2}\epsilon_{(i} {\bar\epsilon}_{j)} ({\rm Im}\epsilon)^{-2}.
\end{equation}
Once a solution is found, to restore a four--dimensional metric one
has merely to solve a linear equation (2) for the rotation one--form $\omega$.
To solve three--dimensional gravity coupled $\sigma$--model 
equations, additional assumptions are needed in general, such as existence 
of the second space--time Killing field commuting with the first one. In 
this case, further two--dimensional reduction leads to completely 
integrable (modified) chiral equations with associated infinite symmetries 
\cite{ge}. Another useful technique consists in restricting a functional
dependence of all target variables on space coordinates through a unique
(several) scalar function \cite{cl}.
However, already information contained in the structure of the target
manifold may be helpful in generating new solutions from already known
ones. This amounts to using the target space isometries to relate
between themselves physically inequivalent field configurations. Then 
it is important to find a concise representation
of symmetry transformations in terms of physical quantities. In
General Relativity it was just the Ernst formulation which
allowed for considerable simplifications. It is natural therefore
to look for similar representation in more general supergravity inspired
bosonic theories.
 
The isometry group of the target manifold is a global symmetry of the 
system, which maps one classical solution to another. For the vacuum
Einstein system (4) it can be written in the $SL(2,R)$ form
\begin{equation} 
\label{z}
\epsilon \rightarrow \frac{a\epsilon+b}{c\epsilon+d},\quad ad-bc=1,
\end{equation} 
with real $a, b, c, d$.
This three--parametric group can be conveniently cast into three 
one--parameter subgroups. The first subgroup is the shift of the complex
Ernst potential on a real constant
\begin{equation} \label{i}
i)\qquad\qquad \epsilon \rightarrow \epsilon+b, \quad (a=d=1,\; c=0).  
\end{equation}
It changes the NUT potential on a constant, what, in view of
(2), does not modify the metric. This transformation is thus a pure gauge.

The second subgroup is the recaling of the Ernst potential
\begin{equation}  \label{ii}
ii) \qquad\qquad \epsilon \rightarrow a^2 \epsilon, \quad (b=c=0,\; d=1/a).  
\end{equation} 
It preserves the r.h.s. of the three--dimensional Einstein equations (7), 
but modifies the four--dimensional metric (1), 
thus producing physically inequivalent field configurations 
(in particular, transforming asymptotically flat solutions into 
asymptotically non--flat ones).

To write down the third subgroup one has first to make a discrete
transformation: an inversion of the Ernst potential 
$\epsilon \rightarrow \epsilon^{-1}$ (what
corresponds to $a=d=0,\; b=-c=1$ in (\ref{z}) plus change of a sign). 
After that one just makes a shift on a real constant:
\begin{equation}   \label{iii}
iii)\quad\quad \epsilon^{-1} \rightarrow \epsilon^{-1} + c,\quad 
(d=a=1,\; b=0). 
\end{equation}
This is the Ehlers transformation \cite{eh}, an essential part of 
the whole group. It is non--linear being 
expressed in terms of $\epsilon$, and physically corresponds to
mixing of a mass and a NUT charge (gravitational analog
of electric--magnetic duality).

The remarkable property of this $\sigma$--model is that
the target manifold is a {\it symmetric} riemannian space. This
property opens prospects to get an infinite--dimensional
symmetry (Geroch group) \cite{ge} making further assumption of the
existence of the second space--time isometry commuting with the first one
(stationary axisymmetric fields, plane waves, non--homogeneous
cosmologies etc.). The corresponding symmetry group will be then
an affine extension of $SL(2, R)$. Some of its lower--level elements
were obtained in various explicit forms as B\"acklund
transformations \cite{bt} and used to generate new exact solutions
of Einstein equations. It is worth noting that the
most sophysticated known exact solutions belong to this type.

Obviously, hidden symmetry of the stationary vacuum Einstein
equations is identical to the $S$--duality \cite{sd}
of the dilaton--axion system (also $SL(2, R)\sim SU(1,1)$), which
is a part of the bosonic sector of $N=4,\, D=4$ supergravity. 
Let us consider first the {\it pure} dilaton--axion system coupled 
to gravity (previous discussion of this model can be found in \cite{bak, mah}. 
Denoting a four--dimensional Peccei--Quinn axion
as $\kappa$ and introducing a complex axidilaton field,
\begin{equation}  
z=\kappa+i{\rm e}^{-2\phi},
\end{equation}
one can write the action as follows
\begin{equation}
S=\int \left\{-R+2\left|\partial z (z-{\bar z})^{-1}\right|^2 \right\}
\sqrt{-g}d^4x.                      
\end{equation}
Clearly the axion--dilaton term has exactly the same symmetries (\ref{z}) 
with $\epsilon$ replaced by $z$. The role of the gauge transformation
now is played by the shift of the axion on a constant, while the inversion
is related to the strong--weak coupling duality transformation.

These symmetries survive upon reduction to three dimensions. 
If the condition of stationarity (1) 
is imposed, the target manifold of the resulting three--dimensional
$\sigma$--model will be the product of two copies
of $SU(1,1)/U(1)$,  one in terms of the Ernst potential, and another
in terms of $z$:
\begin{equation}  
dl^2=2 G_{\alpha {\bar \beta}}dz^\alpha d{\bar z}^\beta=
-2\left\{\frac{d\epsilon d{\bar \epsilon}}{({\bar \epsilon} - \epsilon)^2}
+\frac{dz d{\bar z}}{({\bar z} - z)^2}\right\}.
\end{equation}
Now the K\"ahler metric $G_{\alpha {\bar \beta}},\, (\alpha, \beta=0,1)\,$ 
is generated by the  potential
\begin{equation}  
G_{\alpha {\bar \beta}}=\partial_\alpha \partial_{\bar\beta} 
K(z^\alpha, {\bar z}^\beta), \quad z^\alpha=(\epsilon,\, z),
\end{equation}
\begin{equation}  
K=-\ln V,\quad V={\rm Im} \epsilon \; {\rm Im} z=
f{\rm e}^{-2\phi}.
\end{equation}
The $V$--potential is given by the product of four--dimensional $g_{00}$
and  the dilaton factor playing similar
role in the target space geometry. We observe therefore that the stationary 
dilaton--axion gravity has a remarkable property of a ``square'' of vacuum 
gravity. The reason is simply that the three--dimensional
reductions of both the vacuum Einstein gravity and the dilaton--axion system
have identical $\sigma$--model representations. The coupled system 
also possesses the ``Ernst--axidilaton'' duality symmetry under 
an exchange
\begin{equation}
\epsilon \leftrightarrow z.
\end{equation}

The situation becomes slightly more complicated when vector fields are
included. In three dimensions vector fields can be traded for scalars, and
one can expect to get a larger sigma--model with a higher--dimensional
target manifold \cite{bgm}. This is indeed the case, {\it e.g.},
for the Einstein--Maxwell theory \cite{nk, mg} 
(bosonic sector of $N=2$ supergravity), where one obtains
a K\"ahler target manifold $SU(2,1)/(SU(2)\times U(1))$, as well as for 
other supergravities and dimensionally reduced Kaliza--Klein theories
\cite{bgm}. We will discuss now the $N=4$ supergravity containing a dilaton,
an axion, and six abelian vector fields (for the sake of generality
we take $p$ vector fields). It turns out that the target manifold
is also K\"ahler, and the corresponding complex coordinates are
some generalizations of the Ernst potentials of the Einstein--Maxwell
theory. For the model with only one vector field  
such coordinates were found recently \cite{diak, udu} 
as providing a convenient parametrization for the Ehlers--Harrison 
transformations of this theory discovered earlier \cite{gk} in terms of 
real varables. When several vector fields are present, the target
space is extended rather straightforwardly.

Consider a four--dimensional action
\begin{equation}
S=\int \left\{-R+2\left|\partial z (z-{\bar z})^{-1}\right|^2 +
\left(iz{\cal F}_{\mu\nu}^{n}{\cal F}^{n \mu\nu}+c.c\right)\right\}
\sqrt{-g}d^4x,
\end{equation}
where ${\cal F}^n=(F^n+i{\tilde F}^n)/2,\;
{\tilde F}^{n \mu\nu}=\frac{1}{2}E^{\mu\nu\lambda\tau}F^n_{\lambda\tau}, \,
n=1,...,p$, and the sum over repeated $n$ is understood. For $p=6$ this
is the bosonic sector of $N=4, D=4$ supergravity. This action
is invariant under $SO(p)$ rotations of vector fields, which
is an analog of $T$--duality of dimensionally reduced theories \cite{gpr}.
The equations of motion and Bianchi identities (but not the action) are also 
invariant under $S$--duality transformations
\[
z \rightarrow \frac{az+b}{cz+d},\quad ad-bc=1,
\]
\begin{equation}
F^n \rightarrow (c\kappa+d) F^n
+ c {\rm e}^{-2\phi} {\tilde F}^n.
\end{equation}

Imposing the stationarity condition (1) one can express vector fields 
through the electric $v^n$ and magnetic $u^n$ scalar potentials as follows
\begin{equation}
F^n_{i0}=\frac{1}{\sqrt{2}}\partial_iv^n,
\end{equation}
\begin{equation}
2{\rm Im}\left( z{\cal F}^{n ij}\right)
=\frac{f}{\sqrt{2h}}\epsilon^{ijk}\partial_ku^n.
\end{equation}
In three dimensions the rotation one form $\omega_i$ plays a role of
the graviphoton, and one can show using the standard argument that the
``$T$--duality'' group is enlarged to $SO(1,p+1)$. Also, $S$--duality
becomes the symmetry of the three--dimensional {\it action}. Moreover, 
both these groups turn out to be unified in a larger ``$U$--duality''
group $SO(2,p+2)$ \cite{gk, udu, ht}. 
This can be easily checked by computing the K\"ahler
metric of the resulting target manifold. To find the $\sigma$--model
representation one has to introduce a NUT potential $\chi$  via
\begin{equation}
d\chi=u^n d v^n -v^n d u^n -f^2 *d \omega,
\end{equation}
and to derive the set of equations for $\chi, u^n$ in addition to
the equations for $f, \kappa, \phi, v^n$. The full set of equations
will be that of the three--dimensional gravity coupled $\sigma$--model
possessing the $4+2p$ dimensional target space 
$SO(2,2+p)/\left(SO(2)\times SO(p+2)\right)$. One can parametrize
the target manifold by complex coordinates $z^\alpha,\, \alpha=0,1,...,p+1$
which have the following meaning. The components $\alpha =n=1,...,p$
are complex potentials for vector fields
\begin{equation}
z^n = u^n-z v^n\equiv \Phi^n, \quad n=1,...,p,
\end{equation}
the $\alpha =p+1$ component is the complex axidilaton field itself, 
$z^{p+1}=z$, and  
\begin{equation}
z^0 = \epsilon -  v^n \Phi^n \equiv E,  
\end{equation}
is the $N=4$ analog of the Ernst potential. Somewhat surprisingly, 
the K\"ahler potential, generating the target space metric via (15),
remains untouched by the electric and magnetic potentials and
preserves its value (16) given by the pure dilaton--axion gravity:
\begin{equation}  
K=-\ln V,\quad V={\rm Im} E \;{\rm Im} z +
\left({\rm Im}\Phi^n\right)^2=f{\rm e}^{-2\phi}.
\end{equation}
Hence, in a sense, the ``square'' property of the pure dilaton--axion 
gravity is not destroyed by vectors. At the same time, being expressed
through complex coordinates, K\"ahler potential has non--trivial
dependence on all of them,
so that the metric of the target space is non--degenerate.

Since the K\"ahler metric (5) is given by mixed derivatives upon holomorphic
and antiholomorphic coordinates, a multiplication of $V$ by an arbitrary
holomorphic function and its complex conjugate (to preserve reality of $V$)
does not change the metric. Thus a transformation
\begin{equation}
V(z,\; {\bar z})\rightarrow f(z) {\bar f}({\bar z})V(z,\; {\bar z})
\end{equation}
is the target space isometry. The Ernst--axidilaton duality
(17) (with $\Phi^n$ unchanged) belongs trivially to this class.
Another useful discrete symmetry corresponds to
\begin{equation}
 f(z)=\left(Ez+{\bf \Phi}^2\right)^{-1}, \quad 
 {\bf \Phi}^2 \equiv \Phi^{n\,2},
\end{equation}
and consists in the following:
\begin{equation}
E\rightarrow \frac{z}{Ez+{\bf \Phi}^2},\quad
z\rightarrow \frac{E}{Ez+{\bf \Phi}^2},\quad  
{\bf \Phi}\rightarrow \frac{{\bf \Phi}}{Ez+{\bf \Phi}^2}.  
\end{equation}
Now three--dimensional $U$--duality transformations $SO(2,2+p)$ of $N=4$
supergravity can be listed in the following way. The most obvious symmetries
include $p(p-1)/2\; SO(p)$ rotations acting only on vector fields, 
${\bf \Phi}\rightarrow {\bf \Omega}{\bf \Phi}$, where 
${\bf \Omega}^T{\bf \Omega}=I_p$, as well as $2p+1$ gauge transformations
\begin{eqnarray}
&&{\rm gravitational:}\qquad E\rightarrow E+ g, \qquad 
{\bf \Phi}, z \;{\rm unchanged},\\
&& {\rm magnetic:}\qquad {\bf \Phi}\rightarrow {\bf \Phi}+ {\bf m}, \qquad 
E, z \;{\rm unchanged},\\
&& {\rm electric:}\qquad {\bf \Phi}\rightarrow {\bf \Phi}+ {\bf e}z, \quad 
E\rightarrow E-2{\bf e}{\bf \Phi}-{\bf e}^2 z, \qquad z \;{\rm unchanged},
\end{eqnarray}
and scale
\begin{equation}
E\rightarrow {\rm e}^{2s}E,\quad {\bf \Phi}\rightarrow {\rm e}^s{\bf \Phi},
\quad z \;{\rm unchanged}.
\end{equation}
Here $g,  s, {\bf m}, {\bf e} $ are real scalar and vector group parameters.
The remaining elements of the symmetry group include $2p+1$ Harrison--Ehlers 
transformations, which can be obtained by applying the above discrete maps to
(29--32).  Namely, applying (17) to the electric gauge (31), one gets an 
electric Harrison transformation (the corresponding set of parameters will be 
denoted as ${\bf h}_e$).  Acting by (28) on the magnetic gauge (30) and 
gravitational gauge (29) one obtains a magnetic Harrison (${\bf h}_m$) and 
Ehlers ($c_E$) transformations.
The full group is closed by the $SL(2,R)\; S$--duality (19) expressed in 
terms of the target space variables. This three--parametric set 
can be obtained by applying (17) to gravitational gauge (29), scale (32)
and Ehlers transformation. 

In the particular case $p=1$, due to local isomorphism 
$SL(2,R)\sim Sp(4,R)$, there exists a simple matrix generalization of 
the Ernst potential \cite{diak}. Let us form the $(2\times 2)$ symmetric
complex matrix  collecting K\"ahler coordinates in the following way
\begin{equation}  
{\cal E}=\pmatrix{
E & \Phi \cr
\Phi & -z}.
\end{equation}
One can easily check that the target space metric is reproduced via 
\begin{equation} 
dl^2=-2{\rm Tr}\left\{d{\cal E} \left({\bar {\cal E}}-{\cal E}\right)^{-1}
d{\bar {\cal E}}\left({\bar {\cal E}}-{\cal E}\right)^{-1}\right\},
\end{equation}
what is a direct matrix analog of (4). Also, three--dimensional 
Einstein equations take the form similar to (7):
\begin{equation}  
{\cal R}_{ij}=- 2{\rm Tr}\left\{\left({\bar {\cal E}}-{\cal E}\right)^{-1}
\left(\partial_{(i}{\cal E}\right) \left({\bar {\cal E}}-{\cal E}\right)^{-1} 
\partial_{j)}{\bar{\cal E}}\right\}.
\end{equation}

The analogy with the vacuum General Relativity is suggestive to express 
$U$--duality transformations in a way similar to (9) -- (11) with matrix 
valued parameters. The gauge transformation (9) now is uplifted to
\begin{equation}  
{\cal E} \rightarrow {\cal E}+B,
\end{equation}
where $B$ is the real matrix of parameters 
\begin{equation}  
B=\pmatrix{
g & m \cr
m & b}.
\end{equation}
This matrix--valued gauge transformation joins a gravitational gauge ($g$), 
magnetic gauge ($m$) and an axion shift ($b$) belonging to $S$--duality 
(cf. (9)).

The scale transformation (10) now is splitted into a symmetry
preserving matrix relation:
\begin{equation}  
{\cal E} \rightarrow A^T {\cal E} A.
\end{equation}
Apart from the genuin $SL(2,R)$ scale ($a$), it includes gravitational scale
($s$), electric gauge ($e$) and electric Harrison ($h_e$) transformations: 
\begin{equation}  
A=\pmatrix{
{\rm e}^s & h_e \cr
-e & a}.
\end{equation}

The last subgroup is the linear shift of an inverted matrix
\begin{equation}  
{\cal E}^{-1} \rightarrow {\cal E}^{-1} + C,
\end{equation}
where $C$ is a real symmetric matrix of parameters
\begin{equation}  
C=\pmatrix{
c_E & h_m \cr
h_m & c},
\end{equation}
combining $c$--transformation of $S$--duality with magnetic Harrison 
($h_m$) and Ehlers ($c_E$) transformations.  For pure dilaton--axion
gravity without vector fields the matrices $B, A, C$ become diagonal
and correspond to the product of two $SL(2, R)$. Dilaton--axion gravity
with one vector field generates $Sp(4,R)$ symmetry, as was first found
in \cite{g}. Now to make contact with the $Sp(4,R)$ group, one has merely 
to decompose the matrix Ernst potential into two symmetric real matrices
\cite{diak}
\begin{equation}  
{\cal E}={\cal Q}+ i{\cal P},
\end{equation}
and then construct a $4\times 4$ real matrix
\begin{equation} 
{\cal M}=\left(\begin{array}{crc}
{\cal P}^{-1}&{\cal P}^{-1}{\cal Q}\\
{\cal Q}{\cal P}^{-1}&{\cal P}+{\cal Q}{\cal P}^{-1}{\cal Q}\\
\end{array}\right).
\end{equation}
This is a symmetric symplectic matrix satisfying 
\begin{equation}  
{\cal M}^T J {\cal M} = J,\quad 
J=\pmatrix{
O & I_2 \cr
-I_2 & O}.
\end{equation}
In terms of ${\cal M}$ the metric of the target space reads
\begin{equation}  
dl^2=-\frac{1}{4} {\rm Tr} \{d{\cal M}d{\cal M}^{-1}\},
\end{equation}
while the Einstein equations for $h_{ij}$ are
\begin{equation}  
{\cal R}_{ij}=-\frac{1}{4} {\rm Tr} \{
\left(\partial_{(i} {\cal M}\right)  
\partial_{j)}{\cal M}^{-1} \}.
\end{equation}

Similar decomposition of $U$--duality can be constructed for arbitrary $p$,
but the associated matrix structures are more involved, so we do not 
pursue this here. Rather, consider one another manifestation of 
the ``square'' property of dilaton--axion gravity related to further
two--dimensional reduction. If, in addition to stationarity, an assumption 
of axial symmetry is made (more generally, that of existence of two 
commuting spacetime Killing vectors), the rotation one form in (1) will have
the only non--zero component $\omega_\varphi=\omega$ corresponding to rotation
along the symmetry axis, while the 
three--metric can be written  in the Lewis--Papapetrou gauge
\begin{equation}
h_{ij}dx^i dx^j={\rm e}^{2\gamma}(d\rho^2 + dz^2) +\rho^2 d\varphi^2.
\end{equation}
Then $\gamma$ dissapears from the dynamical equations for the $\sigma$--model variables which now take the form of a modified chiral matrix 
equation \cite{g}
\begin{equation}
\left(\rho {\cal M}_{,\rho} {\cal M}^{-1}\right)_{,\rho}
+\left(\rho {\cal M}_{,z} {\cal M}^{-1}\right)_{,z}.  
\end{equation}
This can serve as a standard input for an application of 
of integrable systems techniques \cite{ge, bt}. A Lax representation
can be written straightforwardly in terms of ${\cal M}$. Vacuum Einstein
theory is a particular case of this system with $\phi=\kappa=v=u=0$.
In that case there exists an alternative chiral equation involving another 
matrix ${\cal F}$ which is expressed directly through $f$ and $\omega$. Since 
$\omega$ and the NUT potential are related non--locally via (2), ${\cal M}$
and ${\cal F}$ representations are essentially different. Meanwhile a 
point--like relation between two pairs $f, \chi$ and $f,\omega$ exists, known 
as Kramer--Neugebauer (KN) map, which transforms ${\cal M}$--equations into
${\cal F}$--equations and vice--versa. This map is particularly helpful
in obtaining the elements of the Geroch group  explicitly. 
The ${\cal F}$--representation for dilaton--axion gravity was found in 
\cite{bdg}:
\begin{equation} 
{\cal F}=\left(\begin{array}{crc}
{\cal P}&-{\cal P}\Omega\\
-\Omega {\cal P}&\Omega{\cal P}\Omega-\rho^2 {\cal P}^{-1}\\
\end{array}\right).
\end{equation}
Here $\Omega$ is a real symmetric matrix
\begin{equation}
\Omega=\pmatrix{
\omega & -q \cr
-q & qv-\beta}, \qquad q=a+v\omega,
\end{equation}
$a=A_{\varphi}$ is the spatial component of the vector potential,
and $\beta=B_{0\varphi}$ is the component of the Kalb--Ramond field (which
was at the origin of $\kappa$). Matrix $\Omega$ generalises $\omega$
of General Relativity, ${\cal P}$ replaces the scalar $f$, while 
${\cal Q}$, entering (43), is a matrix analog of (minus) $\chi$. Similarly
to (two--dimensional reduction of) (2), there exists a non--local relation
between ${\cal Q}$ and $\Omega$ :
\begin{equation}
\nabla {\cal Q}=-\rho^{-1}{\cal P}\left({\tilde \nabla}\Omega\right){\cal P},
\end{equation}
where $\nabla=(\partial_\rho, \partial_z)$ and 
${\tilde\nabla}=(\partial_z, -\partial_\rho)$ are two--dimensional
Hodge dual operators. Now, the {\it local} map between 
${\cal M}$ and ${\cal F}$ is realised by
\begin{equation}
{\cal Q}\rightarrow i\Omega, \qquad  {\cal P}\rightarrow \rho {\cal P}^{-1}.  
\end{equation}
To see this it is sufficient to write down equations for 
${\cal P}$, ${\cal Q}$ and ${\cal P}$, $\Omega$ pairs following from equations 
(48) for ${\cal M}$ and ${\cal F}$ \cite{bdg}. The relation (52) is a 
direct ``matrix square'' of the original KN map
$\chi \rightarrow -i\omega,\;f\rightarrow \rho/f$. Note, that in the both
cases $i$ does not imply complexification, but needed just to
accomodate different signature of cosets relevant to two alternative
representations. Similar KN map exists for arbitrary number $p$
of vector fields. For $p=0$ it was given earlier in \cite{kum}, 
the application of the integrable systems techniques to this case was
recently discussed by Bakas \cite{bak2}.
 
To summarize: coupling of the dilaton--axion system to gravity leads
to a three--dimensional $\sigma$--model with a K\"ahler target manifold 
being a ``square'' of the corresponding General Relativity
manifold. When vector fields are added, the K\"ahler potential 
still preserves its value given by the product of 
Ernst and axidilaton K\"ahler potentials. This gives rise to various
manifestations of the square property of the $N=4$ supergravity
with respect to General Relativity and provides new tools
in the search of clasical solutions to this theory.

\bigskip
 The author wishes to thank the Department of Physics and
Mathematical Physics of the 
Adelaide University for invitation to participate in the First 
Australasian Conference on General Relativity and Gravitation,
for hospitality and support during his visit. Stimulating 
discussions with P. Szekeres, D. Wiltshire, S. Scott,
E. Fackerell and C. Cosgrove are gratefully acknowledged.
Final version of the paper was writen while the author
was visiting the  University of Campinas supported by FAPESP,
he thanks P. Letelier for a stimulating environement and
discussions. This work was supported in part by the RFBR
Grant 96--02--18899. 

\end{document}